\documentclass[superscriptaddress,prd,aps,showpacs,secnumarabic]{revtex4}
\usepackage[dvips]{graphicx,color}
\usepackage{bm}
\usepackage{dcolumn}
\def\st{\mbox{\rule[-5pt]{0pt}{16pt}}}
\def\sb{\mbox{\rule{0pt}{11pt}}}

\def\al{\alpha}

\def\la{\lambda}

\def\ka{\kappa}

\def\ti{\times}
\def\part{\partial}

\newfont{\ex}{cmr10}
\begin{document}
\title{Interference effects and the use of Higgs boson pair production to study the Higgs trilinear self coupling}
\author{Duane A. Dicus}\email{dicus@physics.utexas.edu}\affiliation{Center for Particle Physics, University of Texas, Austin, TX 78712, USA}
\author{Chung Kao}\email{kao@physics.ou.edu}
\affiliation{Homer L. Dodge Department of Physics, University of Oklahoma, Norman, OK 73019, USA }
\author{Wayne W. Repko}\email{repko@pa.msu.edu} \affiliation{Department of Physics and Astronomy, Michigan State University, East Lansing, MI 48824, USA}
\date{\today}
\begin{abstract}
\hspace*{0.1cm} We show that the dominant channel proposed for the determination of the Higgs boson trilinear coupling, $pp\to HH+X$ via gluon fusion, exhibits an interference structure that is independent of the collider energy for collider energies in the range $8\,{\rm TeV}\leq \sqrt{s}\leq 100\,{\rm TeV}$ and is almost maximally destructive. This insensitivity to the collider energy remains approximately true for a variety of other two Higgs production mechanisms although the magnitude of the interference varies widely. 
\end{abstract}
\maketitle
\section{\footnotesize Introduction}
\begin{figure}[h]
\begin{minipage}[t]{0.45\textwidth}
\centering\includegraphics[height=0.75in]{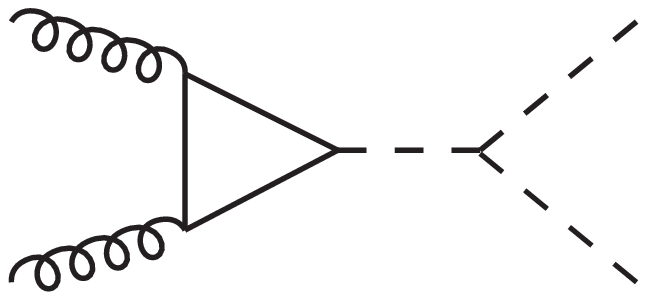}
\end{minipage}%
\begin{minipage}[t]{0.1\textwidth}
\hfil
\end{minipage}%
\begin{minipage}[t]{0.45\textwidth}
\centering\includegraphics[height=0.75in]{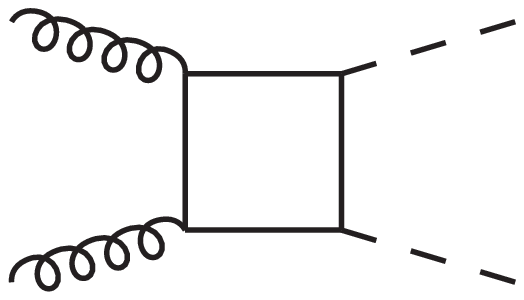}
\end{minipage}%
\caption{The leading order triangle and a representative box diagram are shown.\label{HHdiag}}
\end{figure}

The standard Higgs self interaction is contained in the Higgs potential
\begin{equation}\label{V}
V(H)=\la\,v^2H^2+\la\,vH^3+ \frac{1}{4}\la\,H^4\,,
\end{equation}
where $\la=m_H^2/2v^2$ and, in terms of the $W$ mass, the weak mixing angle and the fine structure constant, $v=M_W\sin\theta_W/\sqrt{\pi\alpha}$. A measurement of the trilinear coupling is an important test of the Standard model behavior of the Higgs-like object discovered at the LHC \cite{ATLAS,CMS}. Given the proliferation of gluon production at the Large Hadron Collider, the natural choice for the study of the Higgs boson trilinear coupling is the gluon fusion process $gg\to HH$ \cite{Bin,Dol,Bag,Goe,Barr,Arh,deFlorian,Hes,Liu,deFl,Fred}. The matrix element comes from the sum of a triangle graph and  several box graphs as indicated in Fig.\,\ref{HHdiag}. Each of these two contributions is gauge invariant. By
calculating the total cross section, $\sigma_{TOT}$ and separately calculating the cross sections from just triangle graph, $\sigma_T$, and just the box graphs, $\sigma_{B}$, we can study the interference between the two amplitudes. Defining interference angle $\alpha_I$ by the equation
\begin{equation}\label{sig}
\sigma_{TOT}\,=\,\sigma_T+\sigma_B+2\,\cos(\alpha_I)\,\sqrt{\sigma_T\sigma_B}\,,
\end{equation}
we can examine how $\cos(\al_I)$ varies as a function of energy. Since the interference term between the box and triangle amplitudes is $2\,\Re e({\cal M}_B{\cal M}_T^*)=2\,|{\cal M}_B||{\cal M}_T|\cos(\al_I)$, $\cos(\al_I)$ is independent of the three Higgs boson self coupling.

\section{\footnotesize Results for $\bm{g}\bm{g}\bm{\to}\bm{H}\bm{H}$}

The numerical results that are shown in Table \ref{Tbl_1} were obtained using $M_H\,=\,125.5$\,GeV, $M_t\,=\,173.1$\,GeV and the MSTW2008 \cite{MSTW} parton distribution functions (PDFs) with both scales equal to the invariant mass of the two Higgs final state. The leading-order amplitudes for Higgs pair production from gluon fusion can be expressed in terms of tensor integrals \cite{pv} and scalar integrals \cite{'thv}. We evaluate these integrals numerically with a FORTRAN code \cite{dk} developed for this purpose. The cross section was calculated in the usual way and then Eq.\,(\ref{sig}) was used to find $\cos(\alpha_I)$. The error in these values of $\cos(\alpha_I)$ from the numerical integration varies from 0.005 at low energies to 0.01 at high energies.
\begin{table}[h]
\begin{center}
\begin{tabular}{|c|c|c|c|c|} \hline
\st\,\,\,$\sqrt{s}\,({\rm TeV})$\,\,\,\,&\,\,\,\,$\sigma_B(fb)$\,\,\,\,&\,\,\,\,$\sigma_T(fb)$ \,\,\,\,&\,\,\,\,$\sigma_{TOT}(fb)$\,\,\,\,&\,\,\,\,\,\,$\cos(\alpha_I)$\,\,\,\,\,\,\\ \hline
\sb 8 & 9.06 & 1.34 & 4.11 & -0.902 \\
13 & 31.6 & 4.36 & 14.9 & -0.898 \\   
14 & 37.8 & 5.16 & 17.9 & -0.898 \\
33 & 243. & 30.3 & 120. & -0.893 \\
60 & 760. & 89.8 & 383. & -0.893 \\
100 & 1900. & 212. & 965. & -0.904 \\ \hline
\end{tabular}
\end{center}
\caption  {For $pp\,\rightarrow\,HH+X$ from $gg\,\rightarrow\,HH$, the center of mass energy, the LO contributions to the cross section from the box, the triangle, and the total cross section, and $\cos(\alpha_I)$. \label{Tbl_1}}
\end{table}
The negative value of $\cos(\al_I)$ reflects the known fact that the interference is destructive. What is surprising is that $\cos(\alpha_I)$ is almost universal.

To check the numbers in Table \ref{Tbl_1}, we multiply the total cross section, $\sigma_{TOT}$, by the $K$ factor given by de Florian and Mazzitelli \cite{deFl}
\begin{equation} \label{K}
K\,=\,1.242\,-\,7.17\left(\frac{E}{1\,{\rm TeV}}\right)^{-1}\,+\,5.77\left(\frac{E}{1\,{\rm TeV}}\right)^{-1/2}\,,
\end{equation}
where $E$ is the center of mass energy. This reproduces their NNLO numbers precisely. 


Since it is not clear how $\cos(\alpha_I)$ can be so constant with energy, it makes sense to ask how it varies with other factors. It is well known that the process, $gg\to HH$ is very sensitive to the values of the scales and the PDFs.  To see how $\cos(\alpha_I)$ depends on the scales and PDF we give in Table \ref{Tbl_2} the box contribution, triangle contribution, total cross section, and $\cos(\alpha_I)$ for three combinations of scale and PDF at three energies.  For each energy the first line is the MSTW PDF with the scales equal to the two Higgs invariant mass (same as Table \ref{Tbl_1}). The second line uses the CTEQ6L1 PDF \cite{CTEQ6L1} with the same scales, and the third line uses the MSTW PDF but with the scales equal to $M_H$. Again the cross sections were calculated from the matrix elements and then $\cos(\alpha_I)$ was determined from Eq.\,(\ref{sig}).
\begin{table}[h]
\begin{center}
\begin{tabular}{|c|c|c|c|c|} \hline
\st\,\,\,$\sqrt{s}\,({\rm TeV})$ \,\,&\,\,\,\,$\sigma_B(fb)$\,\,\,\,&\,\,\,\,\,$\sigma_T(fb)$\,\,\,\,&\,\,\,\, $\sigma_{TOT}(fb)$\,\,\,\,\,&\,\,\,\,$\cos(\alpha_I)$\,\,\,\,\\ \hline
\sb 8 & 9.06 & 1.34 & 4.11 & -0.903  \\
  & 8.05 & 1.20 & 3.64 & -0.902  \\
 &  15.3 & 2.06 & 7.32 & -0.894  \\
14 & 37.8 & 5.16 & 17.9 & -0.897 \\
   & 35.0 & 4.79 & 16.5 & -0.899  \\
   & 60.5 & 7.53 & 29.9 & -0.893  \\ 
100 & 1900. & 212. & 965. & -0.904  \\
    & 1790. & 198. & 914. & -0.902  \\
    & 2530. & 266. & 1340. & -0.887  \\ \hline
\end{tabular}
\end{center}
\caption [Table II] {For the cross sections of $pp\,\rightarrow\,HH+X$ from $gg\,\rightarrow\,HH$ the dependence of $\cos(\alpha_I)$ on the PDF (first and second line at each energy) or on the scale (first and third line at each energy). \label{Tbl_2}}
\end{table}

These results are very striking because, as can be seen, the contributions to the triangle and box cross sections and the total cross section are very dependent on the scales and somewhat dependent on the PDF. But $\cos(\alpha_I)$ apparently doesn't care about such things.

To separate the effects of the underlying physics from that of the parton distribution functions, consider the contributions with the distribution functions set to unity, effectively examining the behavior of the partonic process.  Here we have to consider smaller energies because the triangle graph gets small rapidly due to the s-channel pole. The results are shown in Table \ref{Tbl_3}.

\begin{table}[h]
\begin{center}
\begin{tabular}{|c|c|c|c|c|} \hline
\st\,\,\,$\sqrt{\hat{s}}\,({\rm GeV})$ \,\,&\,\,$\hat{\sigma}_B(fb)$\,\,\,&\,\,\,$\hat{\sigma}_T(fb)$\,\,&\,\,$\hat{\sigma}_{TOT}(fb)$\,\,\,\,&\,\,\,$\cos(\alpha_I)$ \,\,\,\\ \hline
\sb 300 & 0.199 & 0.682$\ti 10^{-1}$ & 0.342$\ti 10^{-1}$ & -1.00 \\ 
400 & 0.964 & 0.875$\ti 10^{-1}$ & 0.482 & -0.980 \\
500 & 0.900 & 0.497$\ti 10^{-1}$ & 0.552 & -0.940 \\
1000 & 0.199 & 0.277$\ti 10^{-2}$ & 0.165 & -0.783 \\
2000 & 0.447$\ti 10^{-1}$ & 0.995$\ti 10^{-4}$ & 0.425$\ti 10^{-1}$ & -0.545 \\
4000 & 0.114$\ti 10^{-1}$ & 0.303$\ti 10^{-5}$ & 0.113$\ti 10^{-1}$ & -0.277 \\ \hline
\end{tabular}
\end{center}
\caption [Table III] {The cross sections and $\cos(\alpha_I)$ with the distribution functions are set to one, i.e., the parton cross sections $\hat{\sigma}(gg\,\to\,HH)$ are shown.\label{Tbl_3}}
\end{table}
The last two lines are uncertain because they depend on the $\sigma_B$ and $\sigma_{TOT}$ being very accurate. However, the conclusion is clear; without the distribution functions to emphasize the low energy parts of the cross sections, $\cos(\alpha_I)$ is no longer universal. More specifically, this Table shows that the behavior of the partonic cross section near threshold ($\sqrt{\hat{s}}\sim 300-500\;{\rm GeV}$) deduced in \cite{livolo,dawson} using unitarity cutting rules is the basically the only region probed by the full cross section $\sigma(pp\to HH+X)$ no matter what the center of mass energy happens to be. 
\section{\footnotesize Interference in Other Two Higgs Processes}
\subsection{Production of two Higgs in association with a {$\bm Z$} or {$\bm W$}}

It is worth looking at other two Higgs production processes to see how $\cos(\alpha_I)$ varies with energy for them. We call the background, $B$, the contribution from the diagrams where both Higgs are emitted separately from the $Z$ line. The signal is the contribution from the diagram with the three Higgs coupling. In analogy to the gluon fusion case above, we call that $T$.  The total cross section is the square of all the diagrams including the cross terms between the $B$ part and the $T$ part. $\cos(\alpha_I)$ is then defined by Eq.\,(\ref{sig}) above.
First consider the results for $pp\,\rightarrow\,ZHH+X$ shown in Table \ref{Tbl_5} for the same energies used in Table \ref{Tbl_1}.  
\begin{table}[h]
\begin{center}
\begin{tabular}{|c|c|c|c|c|}\hline
\st\,\,\,$\sqrt{s}\,({\rm TeV})$ \,\,\,&\,\,\,$\sigma_B$\,\,\,&\,\,\,$\sigma_T$\,\,\,&\,\,\,$\sigma_{TOT}$\,\,\,& \,\,\,$\cos(\alpha_I)$\,\,\,\\ \hline
\sb 8 & 5.37$\ti 10^{-2}$ & 9.10$\ti 10^{-3}$ & 9.38$\ti 10^{-2}$ & 0.701 \\
13 & 0.135 & 2.21$\ti 10^{-2}$ & 0.229 & 0.658 \\
14 & 0.153 & 2.51$\ti 10^{-2}$ & 0.260 & 0.661 \\
33 & 0.588 & 9.31$\ti 10^{-2}$ & 0.975 & 0.628 \\
60 & 1.34  & 0.212 & 2.23  & 0.636 \\
100 & 2.67 & 0.411 & 4.34  & 0.601 \\ \hline
\end{tabular}
\end{center}
\caption[Table V] {Cross sections and interference for $pp\,\rightarrow\,ZHH+X$. The cross sections are in fb.  The different contributions again refer to Eq.\,(\ref{sig}) above. The error in $\cos(\alpha_I)$ due to roundoff is less than or equal to $0.01$. \label{Tbl_5}}
\end{table}

We see that $\cos(\alpha_I)$ changes only from $0.7$ to $0.6$ so unlike gluon production of two Higgs it does change but it doesn't change much. For this reaction and the remaining reactions, the CTEQ6L1 distribution functions were used with scale equal to $\sqrt{\hat{s}}$. We did try varying the distribution functions and found that it makes very little difference which set of PDF's is used. The scale used makes large differences in the values of cross sections but, for a given energy, they all change in the same way so $\cos(\alpha_I)$ is unchanged.   

The processes $pp\,\rightarrow\,W^\pm HH+X$ have an interference behavior similar to $pp\,\rightarrow\,ZHH+X$.  For the same range of energies, $\cos(\alpha_I)$ changes from 0.66 at 8 TeV to 0.57 at 100 TeV for $W^{+}$ and 0.69 to 0.58 for $W^{-}$. The results are shown in Tables \ref{Tbl_6} and \ref{Tbl_7}. 
\begin{table}[h!]
\begin{center}
\begin{tabular}{|c|c|c|c|c|}\hline
\st\,\,\,$\sqrt{s}\,({\rm TeV})$ \,\,&\,\,$\sigma_B$\,\,\,&\,\,\,$\sigma_T$\,\,\,&\,\,\,$\sigma_{TOT}$\,\,\,
&\,\,\,$\cos(\alpha_I)$\,\,\, \\ \hline
\sb 8 & 5.81$\ti 10^{-2}$ & 1.15$\ti 10^{-2}$ & 0.104 & 0.665 \\
13 & 0.136 & 2.61$\ti 10^{-2}$ & 0.237 & 0.629 \\
14 & 0.153 & 2.92$\ti 10^{-2}$ & 0.266 & 0.627 \\
33 & 0.543 & 0.101 & 0.923 & 0.596 \\
60 & 1.21  & 0.221 & 2.03  & 0.579 \\
100 &2.31  & 0.418 & 3.85  & 0.571 \\ \hline
\end{tabular}
\end{center}
\caption[Table VI] {Cross sections and interference for $pp\,\rightarrow\,W^{+}HH+X$. The cross sections are in fb. \label{Tbl_6}}
\end{table}
\begin{table}[h!]
\begin{center}
\begin{tabular}{|c|c|c|c|c|}\hline
\st\,\,\,$\sqrt{s}\,({\rm TeV})$ \,\,\,&\,\,\,$\sigma_B$\,\,\,&\,\,\,$\sigma_T$\,\,\,&\,\,\,$\sigma_{TOT}$\,\,\,
&\,\,\,$\cos(\alpha_I)$\,\,\,\\ \hline
\sb 8 & 2.54$\ti 10^{-2}$ & 5.10$\ti 10^{-3}$ & 4.63$\ti 10^{-2}$ & 0.694 \\
13 & 6.84$\ti 10^{-2}$ & 1.35$\ti 10^{-2}$ & 0.122 & 0.660 \\
14 & 7.86$\ti 10^{-2}$ & 1.54$\ti 10^{-2}$ & 0.139 & 0.647 \\
33 & 0.333 & 6.28$\ti 10^{-2}$ & 0.573 & 0.613 \\
60 & 0.807 & 0.150 & 1.37  & 0.594 \\
100 & 1.64 & 0.300 & 2.76  & 0.585 \\ \hline
\end{tabular}
\end{center}
\caption[Table VII]{Cross sections and interference for $pp\,\rightarrow\,W^{-}HH+X$. The cross sections are in fb. \label{Tbl_7}}
\end{table}

For $p\bar{p}\,\rightarrow\,ZHH$, rather than $pp\,\rightarrow\,ZHH+X$ as given above, $\cos(\alpha_I)$, shown in Table \ref{Tbl_8}, is $0.63$ at 8 TeV and $0.60$ at 100 TeV.  
\begin{table}[h!]
\begin{center}
\begin{tabular}{|c|c|c|c|c|}\hline
\st\,\,\,$\sqrt{s}\,({\rm TeV})$ \,\,&\,\,$\sigma_B$\,\,\,&\,\,\,$\sigma_T$\,\,\,&\,\,\,$\sigma_{TOT}$\,\,\,
&\,\,\,$\cos(\alpha_I)$\,\,\, \\ \hline
\sb 8 & 0.104 & 1.66$\ti 10^{-2}$ & 0.173 & 0.631  \\
14 & 0.217 & 3.40$\ti 10^{-2}$ & 0.360 & 0.634 \\
33 & 0.657 & 0.102 & 1.07  & 0.601  \\
60 & 1.42  & 0.219 & 2.31  & 0.602  \\
100 & 2.72 & 0.417 & 4.41 & 0.598  \\ \hline
\end{tabular}
\end{center}
\caption[Table VIII]{Cross section and interference for $p\bar{p}\,\rightarrow\,ZHH$.  The cross sections are in fb. \label{Tbl_8}}
\end{table}
So the change in the interference with energy when using two quark distributions is even less than when using one quark and one antiquark.
\subsection{Production of two Higgs in association with a {$\bm t\bm\bar{\bm t}$} pair}

Another reaction for producing two Higgs is $pp\,\rightarrow\,t\bar{t}HH+X$. This has contributions from initial quarks and from initial gluons which can not be separated experimentally. But they must be calculated separately so let us first consider the $q\bar{q}\,\rightarrow\,t\bar{t}HH$ part. For all the same input parameters, distribution functions, and scales as above our results are shown in Table \ref{Tbl_9}. Here, $\cos(\alpha_I)$ varies only from $0.84$ to $0.76$ over the usual enormous energy range. The other contribution to $pp\,\rightarrow\,t\bar{t}HH+X$ is from $gg\,\rightarrow\,t\bar{t}HH$.  The results are shown in Table \ref{Tbl_10}.
\begin{table}[h!]
\begin{center}
\begin{tabular}{|c|c|c|c|c|}\hline
\st\,\,\,$\sqrt{s}\,({\rm TeV})$ \,\,\,&\,\,\,$\sigma_B$\,\,\,&\,\,\,$\sigma_T$\,\,\,&\,\,\,$\sigma_{TOT}$\,\,\, &\,\,\,$\cos(\alpha_I)$\,\,\,\\ \hline
\sb 8 & 6.47$\ti 10^{-2}$ & 1.96$\ti 10^{-3}$  & 8.56$\ti 10^{-2}$ & 0.841 \\
13 & 0.212 & 6.54$\ti 10^{-3}$ & 0.280 & 0.825 \\
14 & 0.249 & 7.68$\ti 10^{-3}$ & 0.328 & 0.815 \\
33 & 1.19  & 3.75$\ti 10^{-2}$ & 1.56  & 0.788 \\
60 & 2.98  & 9.51$\ti 10^{-2}$ & 3.89  & 0.765 \\
100 & 6.15 & 0.197 & 8.04  & 0.769 \\ \hline
\end{tabular}
\end{center}
\caption[Table IX]{Contributions to the cross section $pp\,\rightarrow\,t\bar{t}HH+X$ from $q\bar{q}\,\rightarrow\,t\bar{t}HH$. As always the cross sections are in fb. \label{Tbl_9}}
\end{table}
\begin{table}[h!]
\begin{center}
\begin{tabular}{|c|c|c|c|c|}\hline
\st\,\,\,$\sqrt{s}\,({\rm TeV})$ \,\,\,&\,\,\,$\sigma_B$\,\,\,&\,\,\,$\sigma_T$\,\,\,&\,\,\,$\sigma_{TOT}$\,\,\,
&\,\,\,$\cos(\alpha_I$)\,\,\,\\ \hline
\sb 8 & 2.79$\ti 10^{-2}$ & 1.40$\ti 10^{-3}$ & 3.28$\ti 10^{-2}$ & 0.280 \\
13 & 0.179 & 1.03$\ti 10^{-2}$ & 0.211 & 0.253 \\
14 & 0.231 & 1.36$\ti 10^{-2}$ & 0.273 & 0.253 \\
33 & 2.93  & 0.205 & 3.47  & 0.216 \\
60 & 12.6  & 0.966 & 15.0  & 0.206 \\
100 & 38.2 & 3.11  & 45.5  & 0.192 \\ \hline
\end{tabular}
\end{center}
\caption[Table X]{Contributions to the cross section $pp\,\rightarrow\,t\bar{t}HH+X$ from initial gluons. \label{Tbl_10}}
\end{table}
Of course we can't measure the quark and gluon processes separately. If we ignore the fact that the K factors would be different and just add these last two LO processes to get a total cross section for $pp\,\rightarrow\,t\bar{t}HH+X$ we get the result shown in Table \ref{Tbl_11}. 
\begin{table}[h!]
\begin{center}
\begin{tabular}{|c|c|c|c|c|}\hline
\st\,\,\,$\sqrt{s}\,({\rm TeV})$ \,\,\,&\,\,\,$\sigma_B$\,\,\,&\,\,\,$\sigma_T$\,\,\,&\,\,\,$\sigma_{TOT}$\,\,\,
&\,\,\,$\cos(\alpha_I)$\,\,\,\\ \hline
\sb 8 & 9.26$\ti 10^{-2}$  & 3.36$\ti 10^{-3}$  & 0.120  & 0.681 \\
13 & 0.391 & 1.68$\ti 10^{-2}$ & 0.491 & 0.513 \\
14 & 0.480 & 2.12$\ti 10^{-2}$ & 0.600 & 0.490 \\
33 & 4.12  & 0.243 & 5.03  & 0.333 \\
60 & 15.6  & 1.06  & 18.9  & 0.275 \\
100 & 44.4 & 3.31  & 53.6  & 0.243 \\ \hline
\end{tabular}
\end{center}
\caption[Table XI]{Contributions to the cross sections and interference for the total process $pp\,\rightarrow\,t\bar{t}HH+X$. \label{Tbl_11}}
\end{table}
Finally we have a process where $\cos(\alpha_I)$ varies substantially with the energy.
\subsection{{$\bm H\bm H$} production from {$\bm q\bm q\bm\to\bm q\bm q\bm H\bm H$}}
The results for $pp\to qqHH$ via the subprocesses $uu\to uuHH$, $dd\to ddHH$ and $ud\to udHH$ are shown in Tables \ref{Tbl_12}\,,\ref{Tbl_13} and \ref{Tbl_14}. Again, there is very little spread the values of $\cos(\al_I)$ but the interference is, like $gg\to HH$, destructive.
\begin{table}[h!]
\begin{center}
\begin{tabular}{|c|c|c|c|c|}\hline
\st\,\,\,$\sqrt{s}\,({\rm TeV})$ \,\,\,&\,\,\,$\sigma_B$\,\,\,&\,\,\,$\sigma_T$\,\,\,&\,\,\,$\sigma_{TOT}$\,\,\,
&\,\,\,$\cos(\alpha_I)$\,\,\,\\ \hline
\sb 8 & 5.87$\ti 10^{-2}$  & 1.53$\ti 10^{-2}$   & 2.30$\ti 10^{-2}$  & -0.846 \\
13    & 0.161  & 3.81$\ti 10^{-2}$   & 6.90$\ti 10^{-2}$  & -0.830 \\
14    & 0.185  & 4.32$\ti 10^{-2}$   & 8.43$\ti 10^{-2}$  & -0.805 \\
33    & 0.766  & 0.157   & 0.376  & -0.789 \\
60    & 1.80   & 0.341   & 0.938  & -0.768 \\
100   & 3.55   & 0.638   & 1.91   & -0.757 \\ \hline
\end{tabular}
\end{center}
\caption[Table XII]{Contribution to the cross section $pp\to uuHH+X$ from the subprocess $uu\to uuHH$. Again all cross sections are in fb and the errors in $\cos(\alpha_I)$ are less than or equal to $0.01$. \label{Tbl_12}}
\end{table}
\begin{table}[h!]
\begin{center}
\begin{tabular}{|c|c|c|c|c|}\hline
\st\,\,\,$\sqrt{s}\,({\rm TeV})$ \,\,\,&\,\,\,$\sigma_B$\,\,\,&\,\,\,$\sigma_T$\,\,\,&\,\,\,$\sigma_{TOT}$\,\,\,
&\,\,\,$\cos(\alpha_I)$\,\,\,\\ \hline
\sb 8 & 2.10$\ti 10^{-2}$  & 5.80$\ti 10^{-3}$    & 7.80$\ti 10^{-3}$   & -0.861 \\
13    & 6.65$\ti 10^{-2}$   & 1.69$\ti 10^{-2}$    & 2.69$\ti 10^{-2}$   & -0.843 \\
14    & 7.81$\ti 10^{-2}$   & 1.96$\ti 10^{-2}$    & 3.20$\ti 10^{-2}$   & -0.840 \\
33    & 0.404   & 8.95$\ti 10^{-2}$    & 0.186   & -0.809 \\
60    & 1.08    & 0.224    & 0.531   & -0.786 \\
100   & 2.36    & 0.464    & 1.20    & -0.776 \\ \hline
\end{tabular}
\end{center}
\caption[Table XIII]{Contribution to the cross section $pp\to ddHH+X$ from the subprocess $dd\to ddHH$. \label{Tbl_13}}
\end{table}
\begin{table}[h!]
\begin{center}
\begin{tabular}{|c|c|c|c|c|}\hline
\st\,\,\,$\sqrt{s}\,({\rm TeV})$ \,\,\,&\,\,\,$\sigma_B$\,\,\,&\,\,\,$\sigma_T$\,\,\,&\,\,\,$\sigma_{TOT}$\,\,\,
&\,\,\,$\cos(\alpha_I)$\,\,\,\\ \hline
\sb 8 & 0.404  & 0.128    & 0.141   & -0.860 \\
13 & 1.18 & 0.344 & 0.452 & -0.841 \\
14 & 1.36 & 0.394 & 0.530 & -0.836 \\
33 & 6.24 & 1.60  & 2.76  & -0.804 \\
60 & 15.6 & 3.72  & 7.32  & -0.788 \\
100& 32.2 & 7.36  & 15.8  & -0.772 \\ \hline
\end{tabular}
\end{center}
\caption[Table XIV]{Contribution to the cross section $pp\to udHH+X$ from the subprocess $ud\to udHH$. \label{Tbl_14}}
\end{table}
\subsection{{$\bm H\bm H$} production from {$\bm q\bm\bar{\bm q}\bm\to\bm q\bm \bar{\bm q}\bm H\bm H$}}
The contributions from the subprocesses $u\bar{u}\to u\bar{u}HHX$ and $d\bar{d}\to d\bar{d}HH$ are shown in Tables \ref{Tbl_15} and \ref{Tbl_16}.
\begin{table}[h!]
\begin{center}
\begin{tabular}{|c|c|c|c|c|}\hline
\st\,\,\,$\sqrt{s}\,({\rm TeV})$ \,\,\,&\,\,\,$\sigma_B$\,\,\,&\,\,\,$\sigma_T$\,\,\,&\,\,\,$\sigma_{TOT}$\,\,\,
&\,\,\,$\cos(\alpha_I)$\,\,\,\\ \hline
\sb 8 & 1.46$\ti 10^{-2}$ & 3.82$\ti 10^{-3}$ &9.58$\ti 10^{-3}$ & -0.592 \\
13 & 5.03$\ti 10^{-2}$ & 1.27$\ti 10^{-2}$ & 2.92$\ti 10^{-2}$ & -0.669 \\
14 & 6.00$\ti 10^{-2}$ & 1.50$\ti 10^{-2}$ & 3.44$\ti 10^{-2}$ & -0.677 \\
33 & 0.366 & 8.36 $\ti 10^{-2}$ & 0.196 & -0.725 \\
60 & 1.08  & 0.231 & 0.577 & -0.735 \\
100& 2.52  & 0.512 & 1.36  & -0.736 \\ \hline
\end{tabular}
\end{center}
\caption[Table XIV]{Contribution to the cross section $pp\to u\bar{u}HH+X$ from the subprocess $u\bar{u}\to u\bar{u}HH$. \label{Tbl_15}}
\end{table}
\begin{table}[h!]
\begin{center}
\begin{tabular}{|c|c|c|c|c|}\hline
\st\,\,\,$\sqrt{s}\,({\rm TeV})$ \,\,\,&\,\,\,$\sigma_B$\,\,\,&\,\,\,$\sigma_T$\,\,\,&\,\,\,$\sigma_{TOT}$\,\,\,
&\,\,\,$\cos(\alpha_I)$\,\,\,\\ \hline
\sb 8 & 1.33$\ti 10^{-2}$ & 3.56$\ti 10^{-3}$ & 8.95$\ti 10^{-3}$ & -0.575 \\
13 & 4.74$\ti 10^{-2}$ & 1.21$\ti 10^{-2}$ & 2.79$\ti 10^{-2}$ & -0.660 \\
14 & 5.67$\ti 10^{-2}$ & 1.49$\ti 10^{-2}$ & 3.30$\ti 10^{-2}$ & -0.664 \\
33 & 0.359 & 8.30$\ti 10^{-2}$ & 0.193 & -0.721 \\
60 & 1.10  & 0.238 & 0.590 & -0.731 \\
100& 2.61  & 0.538 & 1.41  & -0.733 \\ \hline
\end{tabular}
\end{center}
\caption[Table XVI]{Contribution to the cross section $pp\to d\bar{d}HH+X$ from the subprocess $d\bar{d}\to d\bar{d}HH$. \label{Tbl_16}}
\end{table}
\section{Conclusions}
As discussed in the Introduction, the interference, as parameterized by $\cos(\alpha_I)$, does not depend on the value of the Higgs coupling. For two Higgs production the interference between the graphs with the trilinear coupling and those without is usually constructive. In the exceptional cases of $gg\,\to\,HH$ and $qq\,\to\,qqHH$ the interference is strongly destructive. For all cases the approximate constancy of $\cos(\alpha_I)$ is a result of the near threshold behavior of the amplitudes and the large values of the parton luminosity in the threshold region. For the dominant subprocess, $gg\,\to\,HH$, the almost maximal destructive interference between the triangle and box amplitudes ($\cos(\alpha_I)\,=\,-0.90$) obtained using unitarity cutting \cite{livolo,dawson} tends to make the total cross section small and thus difficult to measure. 

We note that $pp \to t\bar{t} HH +X$ has constructive interference between diagrams with trilinear Higgs couplings and those without,
and the interference angle $\cos(\alpha_I)$ depends on the collider energy as demonstrated in Table X. For collider energy less than 14 TeV, quark-antiquark fusion is the dominant source of $t\bar{t} HH$, while gluon fusion makes
larger contribution for collider energy greater than 33 TeV. Thus parton distribution functions enhance the threshold effect of $q\bar{q} \to t\bar{t} HH$ ($\sqrt{s} \le 14 \; {\rm TeV})$ and $gg \to t\bar{t} HH$ ($\sqrt{s} \ge 33 \; {\rm TeV}$) in $pp \to t\bar{t} HH +X$ at different values of collider energy.

For all the processes, except $gg\to HH$, unitarity arguments are not of use since the leading amplitudes occur at the tree level. Nevertheless, these partonic amplitudes lead to a similar maximal interference in the hadronic cross sections near threshold. As can be seen in Tables XVI and XVII for $u\bar{u}\to ZHH$ and $u\bar{u}\to t\bar{t}HH$, the maximal interference region is not quite as persistent as in the dominant $gg\,\to\,HH$ case so the integration over the relevant parton luminosities results in a clustering of the interference terms rather than yielding a constant value.

\begin{table}[h!]
\centering
\begin{tabular}{|c|c|c|c|c|} \hline
\st\,\,\,$\sqrt{\hat{s}}\,({\rm GeV})$ \,\,&\,\,$\hat{\sigma}_B(fb)$\,\,\,&\,\,\,$\hat{\sigma}_T(fb)$\,\,&\,\,$\hat{\sigma}_{TOT}(fb)$\,\,\,\,&\,\,\,$\cos(\alpha_I)$ \,\,\,\\ \hline
\sb 350 &0.453$\ti 10^{-3}$ & 0.126$\ti 10^{-3}$ & 0.106$\ti 10^{-2}$ & 0.999 \\ 
400  & 0.122$\ti 10^{-1}$ & 0.277$\ti 10^{-2}$ & 0.262$\ti 10^{-1}$ & 0.964 \\
500  & 0.312$\ti 10^{-1}$ & 0.571$\ti 10^{-2}$ & 0.589$\ti 10^{-1}$ & 0.821 \\
1000 & 0.335$\ti 10^{-1}$ & 0.416$\ti 10^{-2}$ & 0.453$\ti 10^{-1}$ & 0.325 \\
2000 & 0.181$\ti 10^{-1}$ & 0.148$\ti 10^{-2}$ & 0.207$\ti 10^{-1}$ & 0.106 \\
4000 & 0.794$\ti 10^{-2}$ & 0.421$\ti 10^{-3}$ & 0.849$\ti 10^{-2}$ & 0.0337 \\ \hline
\end{tabular}
\caption{To separate the effect of the distribution functions we show the partonic cross sections $\hat{\sigma}(u\bar{u}\to ZHH)$ along with $\cos(\alpha_I)$ as a function of $\sqrt{\hat{s}}$.}
\end{table}
\begin{table}[h!]
\centering
\begin{tabular}{|c|c|c|c|c|} \hline
\st\,\,\,$\sqrt{\hat{s}}\,({\rm GeV})$ \,\,&\,\,$\hat{\sigma}_B(fb)$\,\,\,&\,\,\,$\hat{\sigma}_T(fb)$\,\,&\,\,$\hat{\sigma}_{TOT}(fb)$\,\,\,\,&\,\,\,$\cos(\alpha_I)$ \,\,\,\\ \hline
\sb 600 &0.469$\ti 10^{-6}$ & 0.126154$\ti 10^{-7}$ & 0.655$\ti 10^{-6}$ & 1.000 \\ 
700  & 0.340$\ti 10^{-1}$ & 0.101$\ti 10^{-2}$ & 0.464$\ti 10^{-1}$ & 0.975 \\
800  & 0.119              & 0.343$\ti 10^{-2}$ & 0.160              & 0.933 \\
900  & 0.189              & 0.548$\ti 10^{-2}$ & 0.251              & 0.890 \\
1000 & 0.229              & 0.676$\ti 10^{-2}$ & 0.302              & 0.849 \\
1500 & 0.212              & 0.714$\ti 10^{-2}$ & 0.273              & 0.690 \\
2000 & 0.146              & 0.552$\ti 10^{-2}$ & 0.185              & 0.588 \\
\hline
\end{tabular}
\caption{Same as Table XVI for $\hat{\sigma}(u\bar{u}\,\to\,t\bar{t}HH)$.}
\end{table}

\begin{figure}[h]
\begin{minipage}[t]{0.45\textwidth}
\centering\includegraphics[width=3.00in]{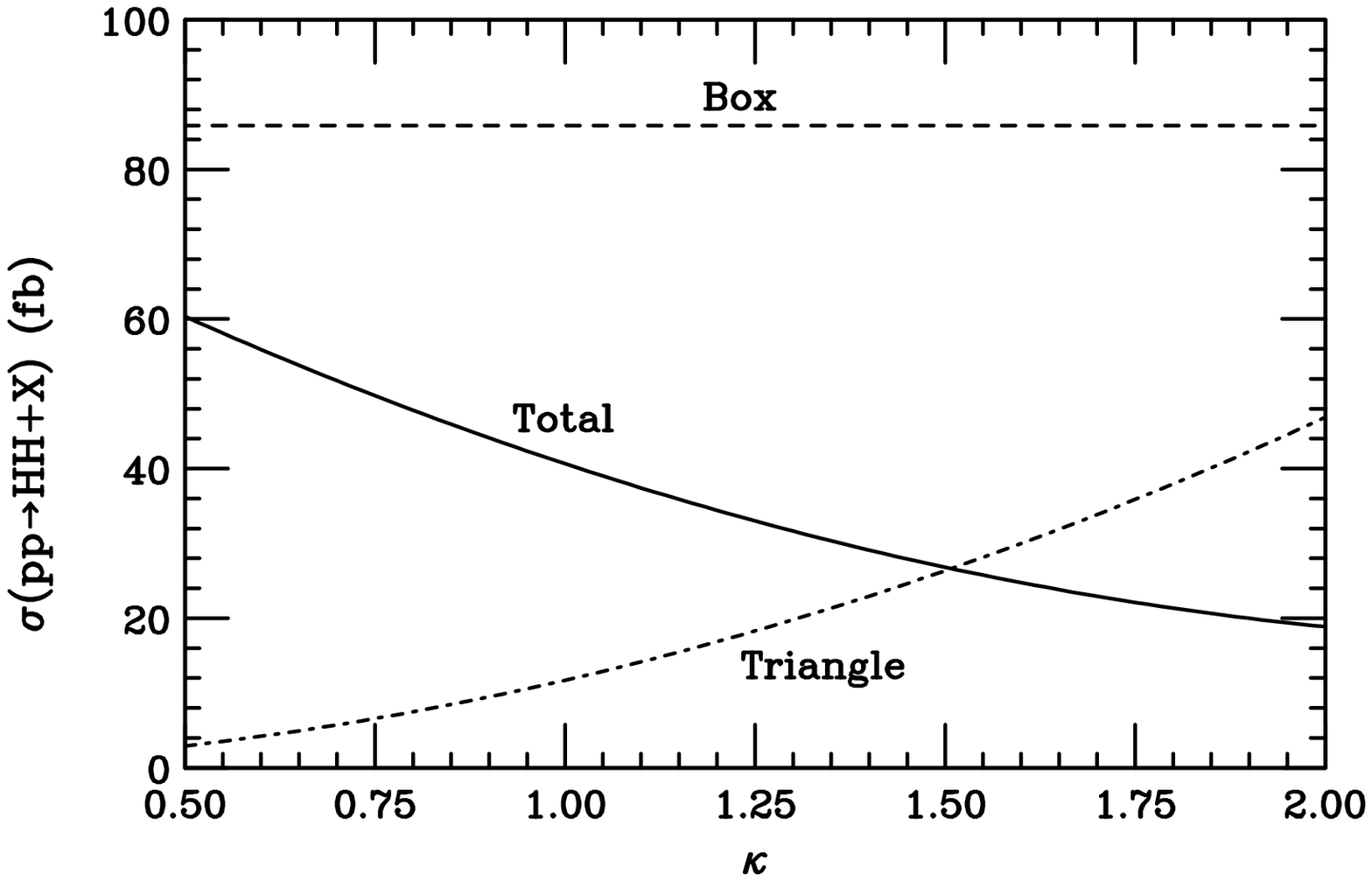}
\caption{The dependence of the $pp\to HH+X$ cross section from the subprocess $gg\to HH$ as a function of the ratio of the trilinear coupling to the Standard model coupling is illustrated for $\sqrt{s}=14$\,TeV and the 2.27 $K$-factor from Eq.\,(\ref{K}). \label{ggfig}}
\end{minipage}%
\begin{minipage}[t]{0.1\textwidth}
\hfil
\end{minipage}%
\begin{minipage}[t]{0.45\textwidth}
\centering\includegraphics[width=3.00in]{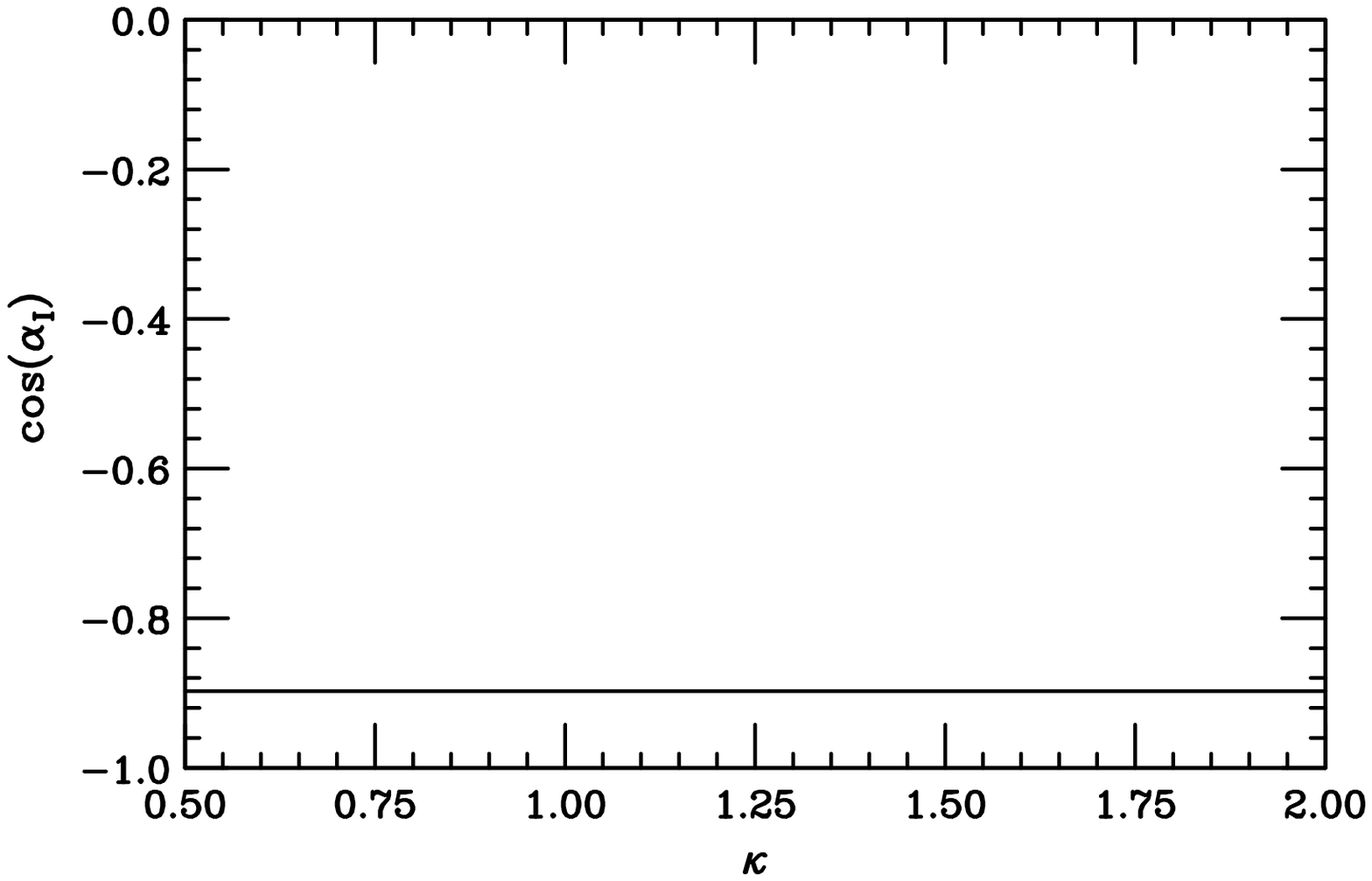}
\caption{The invariance of $\cos(\al_I)$ for changes in the trilinear couplings is shown for the particular case of $\sqrt{s}=14$\,TeV and the 2.27 $K$-factor from Eq.\,(\ref{K}).\label{ggalf}}
\end{minipage}%
\end{figure}

If we allow the standard model Higgs trilinear coupling to vary by a factor $\kappa$ ($\ka\la\,v\,H^3$ in Eq.\,(\ref{V})), the results for $gg\,\to\,HH$ are illustrated in Fig.\,\ref{ggfig}.  In addition, Fig.\,\ref{ggalf} shows that $\cos(\al_I)$ remains constant for any $\kappa$. 

\begin{figure}[b]\centering
\includegraphics[height=3.5in]{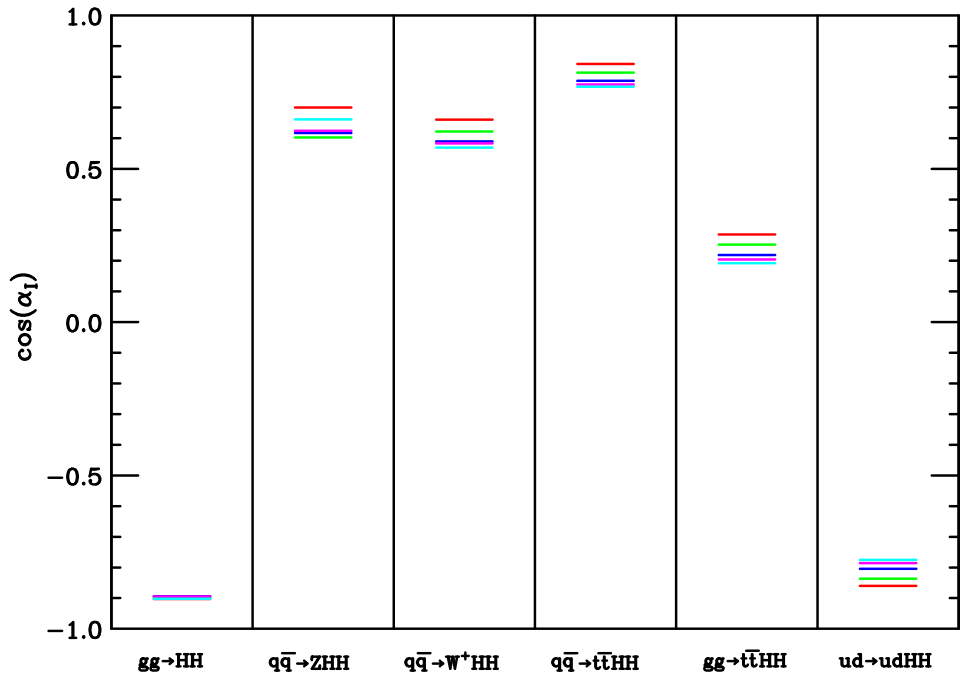}
\caption{\footnotesize The values of $\cos(\al_I)$ are plotted. Note that each partonic initial state has been integrated over the appropriate parton distribution functions for $pp$ collisions. For each process the different lines are for different energies ranging from 8 TeV for the line with the largest magnitude to 100 TeV for the line with the smallest magnitude.} \label{com}
\end{figure}

A summary of the results for $\cos(\al_I)$ given in the Tables is shown in Fig.\ref{com} which illustrates again that, while the value of the interference is very different for different processes, there is very little variation with energy within a given process.
\newpage
\noindent{\bf Acknowledgements}\\
D.~A.~D. was supported in part by the U.~S. Department of Energy under Award No.DE-FG02-12ER41830, C.~K. was supported in part by the U.~S. Department of Energy under Award No.DE-FG02-13ER41979 and W.~W.~R. was supported in part by the National Science Foundation under Grant No. PHY 1068020.

\end{document}